\documentclass[aps,showpacs,twocolumn,prl]{revtex4-1}%

\usepackage{amsmath}
\usepackage{amsfonts}
\usepackage{amssymb}
\usepackage{times}
\usepackage{natbib}%

   \usepackage{graphicx}  
   \DeclareGraphicsExtensions{.eps}

\begin{document}
\title{Breaking the exponential wall in classical simulations of fidelity}
\author{Cesare Mollica}
\author{Tom{\'{a}}{\v{s}} Zimmermann}
\author{Ji\v{r}\'{\i} Van\'{\i}\v{c}ek}
\email{jiri.vanicek@epfl.ch}
\affiliation{Laboratory of Theoretical Physical Chemistry, Institut des Sciences et
Ing\'{e}nierie Chimiques, Ecole Polytechnique F\'{e}d\'{e}rale de Lausanne,
Lausanne, Switzerland}
\date{\today }

\begin{abstract}
We analyze the efficiency of available algorithms for the simulation of
classical fidelity and show that their computational costs increase
exponentially with the number of degrees of freedom for almost all initial
states. Then we present an algorithm whose cost is independent of the system's
dimensionality and show that, within a continuous family of algorithms, our
algorithm is the only one with this property. Simultaneously we propose a
general analytical approach to estimate efficiency of trajectory-based
methods. 
\end{abstract}
\maketitle

\emph{Introduction}. While the solution of the time-dependent Schr\"{o}dinger
equation scales exponentially with dimensionality and is feasible for only a
few continuous degrees of freedom (DOF), classical (CL)\ molecular dynamics
simulations are, in principle, feasible for millions of atoms. It may
therefore be surprising that papers studying classical fidelity
(CF) have provided numerical results for only one or a few DOF
\cite{prosen:2002a,benenti1,eckhardt:2003,*combescure:2007,Karkuszewski:2002,*benenti:2003,*benenti:2003a,*veble:2004,*casati:2005}%
. A notable exception is Ref. \cite{veble:2005}, which, for the largest
systems, relies on initial densities given by characteristic functions. Below
we explain this situation by showing that not only quantum (QM) but also all
previously used CL\ algorithms for fidelity scale exponentially with the
number $D$ of DOF for initial states other than characteristic functions.
Hence even when QM effects are negligible and CL picture is appropriate, the
\textquotedblleft simple\textquotedblright\ CL simulations may be unfeasible.
Since numerical simulations are important for testing analytical theories of
CF in large systems, we design an efficient CF\ algorithm that avoids the
exponential scaling with $D$.

\emph{Quantum and classical fidelity. }While important in its own right, CF
can be viewed as the CL limit of quantum fidelity (QF)
\cite{gorin:2006,*jacquod:2009}, introduced by Peres~\cite{peres:1984} to
measure the stability of QM dynamics (QD). QF is the squared overlap
$F_{\text{QM}}(t)$ at time $t$ of two quantum states, identical at $t=0$, but
evolved with two different Hamiltonians, $H_{0}$ and $H_{\epsilon}%
=H_{0}+\epsilon V$:
\begin{align}
F_{\text{QM}}(t) &  :=\left\vert f_{\text{QM}}(t)\right\vert ^{2},\label{QF}\\
f_{\text{QM}}(t) &  :=\langle\psi\left\vert U_{\epsilon}^{-t}U_{0}%
^{t}\right\vert \psi\rangle,\label{fQM}%
\end{align}
where $f_{\text{QM}}(t)$ is the fidelity amplitude and $U_{\epsilon}^{t}%
:=\exp(-iH_{\epsilon}t/\hbar)$ the QM evolution operator. Rewriting Eq.
(\ref{fQM}) as $f_{\text{QM}}(t)=\langle\psi\left\vert U^{t}\right\vert
\psi\rangle$ with the echo operator $U^{t}:=U_{\epsilon}^{-t}U_{0}^{t}$, it
can be interpreted as the Loschmidt echo, i.e., an overlap of an initial state
with a state evolved for time $t$ with $H_{0}$ and subsequently for time $-t$
with $H_{\epsilon}$. (In general, we write time $t$ as a superscript.
Subscript $\epsilon$ denotes that $H_{\epsilon}$ was used for dynamics. If an
evolution operator, phase space coordinate, or density lacks a subscript,
Loschmidt echo dynamics is implied.) QF amplitude~(\ref{fQM}) is 
ubiquitous in applications: it appears in NMR spin echo experiments
\cite{nmr_echo_4}, neutron scattering~\cite{petitjean:2007}, ultrafast
electronic spectroscopy
\cite{mukamel:1982,*li:1996,*egorov:1998,*shi:2005,*wehrle:2011}, etc. QF
(\ref{QF}) is relevant in QM computation and decoherence
\cite{cucchietti:2003,*gorin:2004}, and can be used to measure nonadiabaticity
\cite{zimmermann:2010a,*zimmermann:2011} or accuracy of molecular QD on an
approximate potential energy surface \cite{li:2009,*zimmermann:2010c}.

Definition (\ref{QF}) can be generalized to mixed states in different ways \cite{gorin:2006,vanicek:2004a,*vanicek:2006}, but we will assume that the initial states are pure. In this case, one may always
write QF (\ref{QF}) as $F_{\text{QM}}(t)=\operatorname{Tr}\left(  \hat{\rho
}_{\epsilon}^{t}\hat{\rho}_{0}^{t}\right)  $ where $\hat{\rho}_{\epsilon}%
^{t}:=U_{\epsilon}^{t}\hat{\rho}U_{\epsilon}^{-t}$ is the density operator at
time $t$. In the phase-space formulation of QM mechanics, QF becomes
{$F_{\text{QM}}(t)={h}^{-{D}}\int dx\rho_{\epsilon,{\text{W}}}^{t}%
(x)\rho_{0,{\text{W}}}^{t}(x)$} where $x:=\left(  q,p\right)  $ is a point in
phase space and {{$A_{\text{W}}(x):=\int d\xi\langle q-\xi/2\left\vert
A\right\vert q+\xi/2\rangle e^{ip\xi/\hbar}$}} is the Wigner transform of $A$.
This alternative form of QF provides a direct connection to its CL limit,
which is precisely the CF, defined as \cite{prosen:2002a, benenti1}
\begin{align}
F_{\text{CL}}(t)  &  :=F_{\text{fid}}(t)=h^{-D}\int dx\rho_{\epsilon}%
^{t}(x)\rho_{0}^{t}(x)\label{CF_fid}\\
&  =F_{\text{echo}}(t)={h}^{-{D}}\int d{x}\rho^{t}(x)\rho^{0}(x)
\label{CF_echo}%
\end{align}
where the first and second line express CF in the fidelity and Loschmidt echo
pictures, respectively, $\rho_{\epsilon}^{t}$ is the CL\ phase-space density
evolved with $H_{\epsilon}$, and $\rho^{t}$ is this density evolved under the
echo dynamics. We omit subscript \textquotedblleft CL\textquotedblright\ for
CL quantities $F$ and $\rho$ since CF is the main subject of this paper.

\emph{Algorithms. }The exponential scaling of QD with $D$ is well known. As
for CF, Eqs.~(\ref{CF_fid})-(\ref{CF_echo}) may be evaluated, e.g., with
trajectory, grid, or mesh-based methods. Clearly, the grid-based methods would
suffer from a similar exponential scaling as QD on a grid. We focus on the
most general and straightforward trajectory-based methods, which are obtained
from Eqs.~(\ref{CF_fid})-(\ref{CF_echo}) using the Liouville theorem, yielding
equivalent expressions
\begin{align}
F_{\text{fid}}(t)  &  ={h}^{-{D}}\int d{x^{0}}\rho(x_{\epsilon}^{-t}%
)\rho(x_{0}^{-t})\text{ \ and}\label{CF_fid_tr}\\
F_{\text{echo}}(t)  &  ={h}^{-{D}}\int d{x^{0}}\rho(x^{-t})\rho(x^{0}).
\label{CF_echo_tr}%
\end{align}
Above, $x_{\epsilon}^{t}:=\Phi_{\epsilon}^{t}(x^{0})$ where $\Phi_{\epsilon
}^{t}$ is the Hamiltonian flow of $H_{\epsilon}$ and $x^{t}:=\Phi^{t}(x^{0})$
where $\Phi^{t}:=\Phi_{\epsilon}^{-t}\circ\Phi_{0}^{t}$ is the Loschmidt echo
flow. Since it is the phase space points rather than the densities that evolve
in expressions (\ref{CF_fid_tr})-(\ref{CF_echo_tr}), we can take $\rho
=\rho_{\text{W}}$, i.e., the Wigner transform of the initial QM state. We
further rewrite Eqs.~(\ref{CF_fid_tr})-(\ref{CF_echo_tr}) in a form suitable
for Monte Carlo evaluation, i.e., as an average
\[
\left\langle A(x^{0},t)\right\rangle _{W(x^{0})}:=\frac{\int d{x^{0}}%
A(x^{0},t)W(x^{0})}{\int d{x^{0}}W(x^{0})}%
\]
where $W$ is the sampling weight for initial conditions $x^{0}$. The weight
can be any positive definite function, but it is advantageous to consider the
weight to be related to the density $\rho$. While previously used algorithms
sampled from $\rho$
\cite{benenti1,Karkuszewski:2002,*benenti:2003,*benenti:2003a,*veble:2004,*casati:2005,veble:2005}%
, we consider more general weights $W=W_{M}(x^{0}):=\rho(x^{0})^{M}$ and
$W=W_{M}(x_{0}^{-t})=\rho(\Phi_{0}^{-t}(x^{0}))^{M}$ for the echo and fidelity
dynamics, respectively. These weights yield $M$-dependent algorithms%
\begin{align}
F_{\text{fid-}M}(t)  &  =I_{M}{\langle\rho(x_{\epsilon}^{-t})\rho(}x_{0}%
^{-t})^{1-M}{\rangle}_{\rho(x_{0}^{-t})^{M}},\label{CF_fid_M}\\
F_{\text{echo-}M}(t)  &  =I_{M}{\langle\rho(x^{-t})\rho(}x^{0})^{1-M}{\rangle
}_{\rho(x^{0})^{M}}, \label{CF_echo_M}%
\end{align}
where $I_{M}:=h^{-D}\int\rho(x^{0})^{M}dx^{0}$ is a normalization factor. In
both families of algorithms (\ref{CF_fid_M})-(\ref{CF_echo_M}), sampling can
be done by Metropolis Monte Carlo for general dynamics and any positive
definite weight $\rho^{M}$. For $M>0$, the echo algorithms (\ref{CF_echo_M})
are, however, much more practical since the initial state is often known explicitly (and generally is much smoother than the final state), making sampling easier. Furthermore, for simple
initial states such as Gaussian wavepackets (GWPs), the Metropolis sampling in
the echo algorithms can be replaced by analytical sampling. Therefore, for
$M>0$ the fidelity algorithms are more of a theoretical possibility than a
practical tool. For $M=0$, the sampling is uniform and makes sense only for a
compact phase space of finite volume $\Omega=\Omega_{1}^{D}=\left(
n_{1}h\right)  ^{D}$ where $\Omega_{1}$ and $n_{1}$ are respectively the phase
space volume and Hilbert-space dimension for a single DOF. For $M>0$,
importance sampling based on the weight $W_{M}$ is used and an infinite phase
space is allowed. For general $M$, the sampling is only defined for CL states
(such as GWPs), for which $\rho\geq0$. However, for $M=0$ and for the
important special case of $M=2$, the sampling is defined for any pure state,
i.e., even for negative values of $\rho$.

In order to compute CF directly from algorithms (\ref{CF_fid_M}) or
(\ref{CF_echo_M}), the normalization factor $I_{M}$ must be known
analytically. For general pure states, $I_{M}$ is known analytically only for
$M=0,1$, or $2.$ For $M=0$, $I_{0}=n_{1}^{D}$ because of the requirement of
finite phase space. For both $M=1$ and $M=2$, $I_{M}=1$ since
$\operatorname{Tr}\hat{\rho}=\operatorname{Tr}\hat{\rho}^{2}=1$. For
$M\notin\{0,1,2\}$, algorithms (\ref{CF_fid_M}) and (\ref{CF_echo_M}) can only
be used for special initial states. E.g., for initial GWPs $\rho
(x)={g}(x;X,a):=2^{D}\exp\left[  -(q-Q)^{2}/a^{2}-(p-P)^{2}a^{2}/\hbar
^{2}\right]  $ where $X$ is the center and $a$ the width of the GWP, we have
$I_{M}=\left(  2^{M-1}/M\right)  ^{D}$ for general $M>0$. However, the unknown
normalization factor can be removed from Eqs. (\ref{CF_fid_M}) and
(\ref{CF_echo_M}) by dividing them by the value of $I_{2}$ [note that
$I_{2}(0)=F(0)$] obtained with the same algorithm and trajectories. Resulting
\textquotedblleft normalized\textquotedblright\ (N)\ algorithms,
\begin{align}
F_{\text{fid-N-}M}(t)  &  :=\frac{F_{\text{fid-}M}(t)}{I_{2}(t)}%
=\frac{{\langle\rho(x_{\epsilon}^{-t})\rho(}x_{0}^{-t})^{1-M}{\rangle}%
_{\rho(x_{0}^{-t})^{M}}}{{\langle}\rho(x_{0}^{-t})^{2-M}{\rangle}_{\rho
(x_{0}^{-t})^{M}}},\label{CF_fid-N_M}\\
F_{\text{echo-N-}M}(t)  &  :=\frac{F_{\text{echo-}M}(t)}{I_{2}(0)}%
=\frac{{\langle\rho(x^{-t})\rho(}x^{0})^{1-M}{\rangle}_{\rho(x^{0})^{M}}%
}{{\langle\rho(}x^{0})^{2-M}{\rangle}_{\rho(x^{0})^{M}}}, \label{CF_echo-N_M}%
\end{align}
are practical for general initial states and for any $M$. As far as we know,
from the four families of algorithms (\ref{CF_fid_M}), (\ref{CF_echo_M}),
(\ref{CF_fid-N_M}), and (\ref{CF_echo-N_M}) only echo-1 (\ref{CF_echo_M}) has
been used
previously~\cite{benenti1,Karkuszewski:2002,*benenti:2003,*benenti:2003a,*veble:2004,*casati:2005,veble:2005}%
. Note however, that for initial states given by characteristic functions,
echo-1 = echo-$M$ = echo-N-$M$ for all $M>0$.

\emph{Efficiency. }The cost of a typical method propagating $N$ trajectories
for time $t$ is $O(c_{\text{f}}tN)$ where $c_{\text{f}}$ is the cost of a
single force evaluation. However, among the above mentioned algorithms, this
is only true for the fidelity algorithms with $M=0$. Remarkably, in all other
cases, the cost is $O(c_{\text{f}}t^{2}N)$. For a single time $t$, the cost is
linear in time, but if one wants to know CF for all times up to $t$, the cost
is quadratic with $t$. For the echo algorithms, it is because one must make
full backward propagation for each time between $0$ and $t$. For the fidelity
algorithms, it is because the weight function $\rho(x^{-t})^{M}$ changes with
time and the sampling has to be redone from scratch for each time between $0$
and $t$. In other words, different trajectories are used for each time between
$0$ and $t$.

The above estimates are correct but not the full story. There are hidden
costs\ since the number of trajectories $N$ required for convergence can
depend on $D$, $t$, dynamics, initial state, and method. One usually empirically
increases $N$ until convergence, but this is often impracticable. Instead, we
estimate $N$ analytically. An essential point is that $N$ is fully determined
by the desired discretization error $\sigma_{\text{discr}}$. The expected
systematic component of $\sigma_{\text{discr}}$ is zero or $O(N^{-1})$ for all
cases studied and is negligible to the expected statistical component
$\sigma=O(N^{-1/2})$ which therefore determines convergence.\ This statistical
error is computed as $\sigma^{2}(t,N)={\overline{F(t,N)^{2}}-\overline
{F(t,N)}^{2}}$ where the overline denotes an average over infinitely many
independent simulations with $N$ trajectories. Hence we can formulate the
problem of efficiency precisely: \textquotedblleft What $N$ is required to
converge fidelity $F$ to within a statistical error $\sigma$%
?\textquotedblright\ We let $N$ be a function of $F$ because in many
applications, one is interested in $F$ above a certain threshold value
$F_{\text{min}}$. This threshold can vary with application: it may be close to
unity (in quantum computing) or to zero (yet finite, in calculations of
spectra), but in general will be \emph{independent} of $D$.

The discretized form of Eq.~(\ref{CF_fid_M}) is $F_{\text{fid-}M}%
(t,N)=I_{M}N^{-1}\sum_{j=1}^{N}\rho_{\text{CL}}(x_{\epsilon,j}^{-t}%
)\rho_{\text{CL}}(x_{0,j}^{-t})^{1-M}$, from which $\overline{F_{\text{fid-}%
M}(t,N)^{2}}=I_{M}^{2}N^{-1}\langle\rho(x_{\epsilon}^{-t})^{2}\rho(x_{0}%
^{-t})^{2-2M}\rangle_{\rho(x_{0}^{-t})^{M}}+(1-N^{-1})F^{2}$. Similarly, from
Eq. (\ref{CF_echo_M}) $F_{\text{echo-}M}(t,N)=I_{M}N^{-1}\sum_{j=1}^{N}%
\rho(x_{j}^{-t})\rho(x_{j}^{0})^{1-M}$, hence $\overline{F_{\text{echo-}%
M}(t,N)^{2}}=I_{M}^{2}N^{-1}\langle\rho(x^{-t})^{2}\rho(x^{0})^{2-2M}%
\rangle_{\rho(x^{0})^{M}}+(1-N^{-1})F^{2}$.

Realizing that $\overline{F_{\text{fid-}M}(t,N)}=\overline{F_{\text{echo-}%
M}(t,N)}=F(t)$ in both cases, we obtain the same error
\begin{align}
\sigma_{\text{fid-}M}^{2}  &  =\sigma_{\text{echo-}M}^{2}=N^{-1}(I_{M}%
J_{M}-F^{2}),\label{stat-M-theor}\\
J_{M}  &  :=h^{-D}\int dx^{0}\rho(x^{-t})^{2}\rho(x^{0})^{2-M}. \label{J_M}%
\end{align}
In the special case of $M=2$, we find our\emph{ main result},
\begin{equation}
\sigma_{\text{fid-}2}^{2}=\sigma_{\text{echo-}2}^{2}=N^{-1}\left(
1-F^{2}\right)  . \label{stat-2}%
\end{equation}
This expression shows that for general states and for general dynamics,
statistical error of $F_{\text{fid-}2}$ or of $F_{\text{echo-}2}$ depends only
on $N$ and $F$. In other words, the number of trajectories needed for
convergence is\emph{\ independent} of $t$, $D$, or dynamics of the system.
This important result is due to the fact that for the sampling weight
$W=\rho^{2}$, each numerical trajectory contributes evenly to the weighted average (at time $t=0$).

As for algorithms (\ref{CF_fid_M})-(\ref{CF_echo_M}) with $M\neq2$, one might
hope to improve convergence by employing the normalized versions
(\ref{CF_fid-N_M})-(\ref{CF_echo-N_M}). The error analysis is simplified using
the formula for statistical error of a ratio of two random variables,%
\begin{equation}
\left(  \frac{\sigma_{A/B}}{\overline{A/B}}\right)  ^{2}=\left(  \frac
{\sigma_{A}}{\bar{A}}\right)  ^{2}+\left(  \frac{\sigma_{B}}{\bar{B}}\right)
^{2}-2\frac{\overline{AB}-\bar{A}\bar{B}}{\bar{A}\bar{B}}. \label{prop_err}%
\end{equation}
In our case, $F_{\text{N-}M}(t,N)=A/B$ where $A=F_{M}(t,N)$, $B=F_{M}(0,N)$,
$\bar{A}=F(t)$, $\bar{B}=F(0)=1$, and $\sigma_{A}$ and $\sigma_{B}$ are given
by Eq. (\ref{stat-M-theor}). The only unknown in Eq. (\ref{prop_err}) is
$\overline{AB}$. For the normalized echo algorithms (\ref{CF_echo-N_M}), we
have $\overline{AB}=\overline{F_{\text{echo-}M}(t,N)F_{\text{echo-}M}%
(0,N)}=I_{M}^{2}N^{-1}\langle\rho(x^{-t})\rho(x^{0})^{3-2M}\rangle_{\rho
(x^{0})^{M}}+(1-N^{-1})F(t)F(0)=N^{-1}I_{M}K_{M}+(1-N^{-1})F$ where
$K_{M}:=h^{-D}\int dx^{0}\,\rho(x^{-t})\rho(x^{0})^{3-M}$. The same derivation
goes through for the fidelity algorithms. The final error can in both cases be
written as%
\begin{equation}
\sigma_{\text{N-}M}^{2}=N^{-1}\left(  J_{M}-2K_{M}F+I_{4-M}F^{2}\right)  .
\label{stat-N-M-theor}%
\end{equation}

\emph{Exponential growth of the error for} $M\neq2$.\emph{ }Now we will show
that the special case $M=2$ is unique and that all the other above-mentioned
algorithms (which include all the algorithms available in the literature) have
an error growing exponentially with $D$. Since we are searching for
counterexamples, special cases are sufficient. For us these will be initial
GWP states and \textquotedblleft pure displacement\textquotedblright\ (PD) or
\textquotedblleft pure squeezing\textquotedblright\ (PS) dynamics
\cite{eckhardt:2003,*combescure:2007}. All calculations can be done
analytically using special cases of the integral%
\begin{align*}
&  \int dq\exp[-c_{1}(q-q_{1})^{2}-c_{2}(q-q_{2})^{2}]\\
&  =\left(  \frac{\pi}{c_{1}+c_{2}}\right)  ^{D/2}\exp\left[  -\frac
{c_{1}c_{2}}{c_{1}+c_{2}}\left(  q_{1}-q_{2}\right)  ^{2}\right]  .
\end{align*}

In the PD case, the center of the GWP moves while both its shape and size
remain constant. Such fidelity dynamics can be realized exactly by two
displaced simple harmonic oscillator (SHO) potentials with equal force
constants. For PD, the width $a_{0}^{t}=a_{\epsilon}^{t}=a^{t}=a^{0}=a$ and
either $X_{\epsilon}^{t}=X_{0}^{t}+\Delta X^{t}$ or $X^{t}=X^{0}+\Delta X^{t}%
$. CF is {$F(t)=h^{-D}\int dx\,g(x;X^{t},a)g(x;X^{0},a)={\exp}\left\{
-\frac{1}{2}\left[  \left(  \frac{{\Delta Q}^{t}}{a}\right)  ^{2}+\left(
\frac{{\Delta P}^{t}a}{\hbar}\right)  ^{2}\right]  \right\}  $} and the factor
(\ref{J_M}) needed in the statistical error can be expressed in terms of $F$
as $J_{M}=\left(  \frac{2^{3-M}}{4-M}\right)  ^{D}F^{\gamma_{M}}$ with
$\gamma_{M}=4-8/(4-M).$ Using this result in Eq. (\ref{stat-M-theor}), the
statistical errors are%
\begin{align}
\sigma_{\text{fid-}M\text{, PD}}^{2}  &  =\sigma_{\text{echo-}M\text{, PD}%
}^{2}=\frac{1}{N}\left(  \beta_{M}^{D}F^{\gamma_{M}}-{F^{2}}\right)
{,}\label{stat-M-analyt-PD}\\
\beta_{0}  &  =2n_{1}\text{ and }\beta_{0<M<4}=\frac{4}{(4-M)M}.
\label{beta_M-PD}%
\end{align}
Note that $\beta_{M}\geq1$ and the minimum $\beta_{2}=1$ is achieved for
$M=2$. The minimum agrees precisely with the general result (\ref{stat-2}).
Except for $M=2$, $\beta_{M}>1$, showing that even in the simple case of PD
dynamics, the errors of all algorithms from the families (\ref{CF_fid_M}) and
(\ref{CF_echo_M}) grow \emph{exponentially} with $D$, which is the
\emph{second major result} of this paper. The normalized methods
(\ref{CF_fid-N_M}) and (\ref{CF_echo-N_M}) lower the prefactor of the error
but do not change the exponential scaling with $D$: Since $K_{M}%
=[2^{3-M}/(4-M)]^{D}F^{\delta_{M}-1}$ where $\delta_{M}=3-2/(4-M)$,
statistical errors are
\[
\sigma_{\text{fid-N-}M\text{, PD}}^{2}=\sigma_{\text{echo-N-}M\text{, PD}}%
^{2}=N^{-1}\beta_{M}^{D}\left(  F^{\gamma_{M}}+F^{2}-2F^{\delta_{M}}\right)
.
\]

In the PS case, the center of the GWP remains fixed while its width narrows in
some directions and spreads in others. Such fidelity dynamics is realized
exactly by two inverted SHOs with common centers and different force
constants. Analytical calculations show that the errors of different
algorithms again grow \emph{exponentially} with $D$ (see Table I).

To summarize, in all cases studied, for $D\gg1$ the number of trajectories
required for a specified convergence is%
\begin{equation}
{N=\sigma^{-2}\alpha(F}){\beta^{D}} \label{N_vs_D}%
\end{equation}
where $\alpha$ and $\beta${\ depend on the method and dynamics and are listed
}in Table~\ref{Costs}. For both fidelity and echo algorithms with $M=2$, for
any dynamics and any initial state, the coefficient $\beta=1$, implying
independence of $D.$ Note also that algorithms with $M=2$ are automatically
normalized. For all other algorithms (both echo and fidelity, both
unnormalized and normalized, and for any $M\neq2$) and for both PD and PS
dynamics, $\beta>1$, implying an exponential growth with $D$. This growth is
dramatic for $M=0$ ($\beta=2n_{1}\gg1$): since $n_{1}^{D}$ is the Hilbert
space dimension, the cost of $M=0$ algorithms approaches that of QF. This is
unfortunate since $F_{\text{fid-}0}$ is the only algorithm that scales
linearly in time. On the other hand, for the most intuitive and most common
$M=1$ algorithms, $\beta=4/3$ or $\sqrt{2}$, and the growth is much slower,
although still exponential. We cannot exclude existence of a faster CL
algorithm; however, we doubt existence of a CL algorithm that would be
\emph{both} linear in $t$ \emph{and} independent of $D$.

\begin{table}
[th]\begin{ruledtabular}
\begin{tabular}{lllll}
Method & Dynamics type& $\alpha(F)$ & $\beta$ \tabularnewline
\hline
fid-0 & displacement & $F^2$ & $2n_1$ \\
fid-0 & squeezing & $F$ & $2n_1$ \\
echo-1 & displacement & $F^{4/3}$ & $4/3$ \\
echo-1 & squeezing, $F\approx 1$ & $1$ & $4/3$ \\
echo-1 & squeezing, $F\ll 1$ & $F$ & $\sqrt{2}$\\
echo-1' & displacement & $1-F^{4/3}$ & $4/3$ \\
echo-1' & squeezing, $F\approx 1$ & $\frac{8}{9} (1-F)$ & $4/3$ \\
echo-1' & squeezing, $F\ll \left(\frac{8}{9}\right)^{\frac{D}{2}}$ & $1$ & $4/3$\\
echo-N-1 & displacement & $ F^{2} + F^{4/3} - 2 F^{7/3}$ & $4/3$ \\
echo-N-1 & squeezing, $F\approx 1$ & $\frac{8}{9} (1-F)$ & $4/3$ \\
echo-N-1 & squeezing, $F\ll 1$ & $F$ & $\sqrt{2}$\\
echo-2 & general, general state & $1 - F^2$ & $1$ \\
\end{tabular}
\caption{\label{Costs} The number of trajectories needed to achieve a given $\sigma$
is for $D\gg 1$ given by $N = \sigma^{-2} \alpha (F) \beta ^{D}$.
The table lists $\alpha (F)$ and $\beta$ for different cases. Note that fid-0, echo-1, echo-1', and echo-N-1 results are
for initial GWPs and exhibit exponential scaling with $D$ while echo-2 result, valid for any
initial state, is independent of $D$.}
\end{ruledtabular}

\end{table}

\emph{Numerical results and conclusion.}To illustrate the analytical results
obtained above, numerical tests were performed in multidimensional systems of
uncoupled displaced SHOs (for PD dynamics), inverted SHOs (for PS dynamics),
and perturbed kicked rotators (for nonlinear integrable and chaotic dynamics).
The last model is defined, $\operatorname{mod}${$(2\pi)$}, by the map
{$q_{j+1}=~q_{j}+p_{j}$}, {$p_{j+1}=~p_{j}-\nabla W(q_{j+1})-\epsilon\nabla
V(q_{j+1})$} where {$W(q)=-k\cos q$} is the potential and {$V(q)=-\cos(2q)$}
the perturbation of the system; $k$ and $\epsilon$ determine the type of
dynamics and perturbation strength, respectively. Uncoupled systems were used
in order to make QF calculations feasible (as a product of $D$ 1-dimensional
calculations); however, the CF calculations were performed as for a truly
$D$-dimensional system. The initial state was always a multidimensional GWP.
Expected statistical errors were estimated by averaging actual statistical
errors over $100$ different sets of $N$ trajectories. No fitting was used in
any of the figures, yet all numerical results agree with the analytical
estimates. Note that Table I and figures show results for algorithm echo-1',%
\[
F_{\text{echo-1'}}(t)=1+{\langle\rho(x^{-t})-\rho(x^{0})\rangle}_{\rho(x^{0}%
)},
\]
which is a variant of echo-1 accurate for high fidelity. Both echo-1 and
echo-1' reduce to echo-N-1 if normalized.

\begin{figure}
[hptb]\centerline{\resizebox{\hsize}{!}{\includegraphics[]{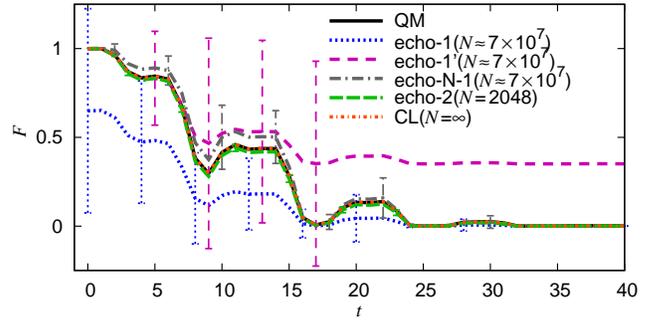}}}
\caption{\label{fidelity_100D}Convergence of different classical fidelity algorithms in a $100$-dimensional system of
perturbed~($\epsilon= 10^{-4}$)
quasi-integrable~($k=0.2$) kicked rotators with $n_{1}=131072$.
Algorithm echo-2 agrees with the QM result and
converges with only $N=2048$ trajectories whereas the echo-1, echo-1', and echo-N-1 results are
far from converged even with $N\approx 7 \times 10^{7}$. Fully converged CL$(N = \infty )$ is computed as a
product of 100 one-dimensional fidelities. The ``hopelessly'' unconverged fid-0 algorithm not shown.
For clarity, echo-1' error bars not shown for $t>20$.}
\end{figure}

\begin{figure}
[hptb]\centerline{\resizebox{\hsize}{!}{\includegraphics[]{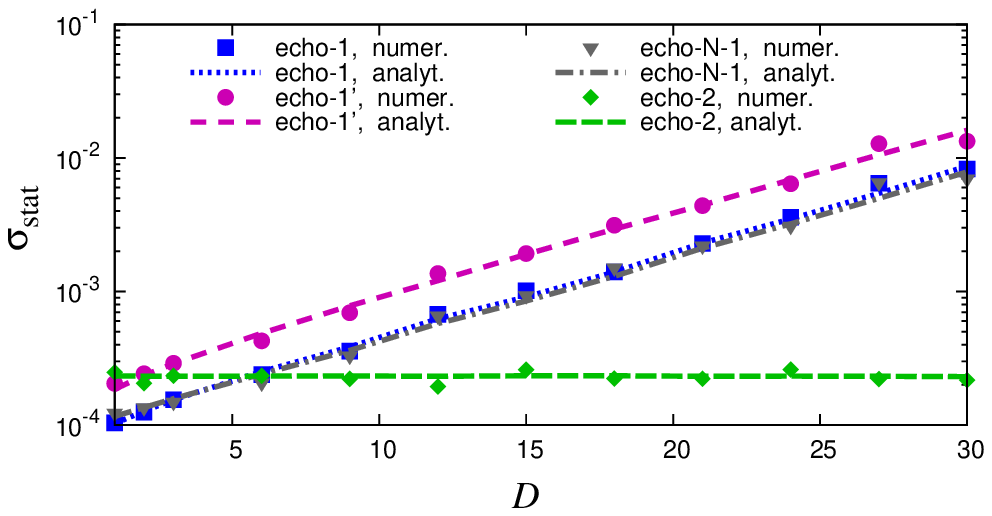}}}
\caption{\label{stat_err_PD}Statistical error grows exponentially with $D$ for the echo-1, echo-1', and echo-N-1
algorithms and is independent of $D$ for the echo-2 algorithm. Dynamics corresponds to pure displacement, $N \approx 10^7$,
and time was chosen separately for each $D$ so that $F\approx0.3$.}
\end{figure}

\begin{figure}
[hptb]\centerline{\resizebox{\hsize}{!}{\includegraphics[]{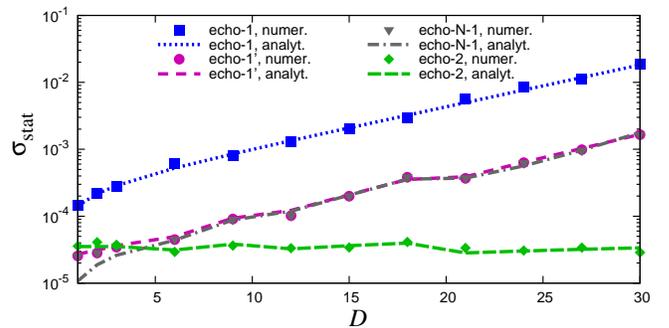}}}
\caption{\label{stat_err_PS}Statistical error grows exponentially with $D$ for the echo-1, echo-1', and echo-N-1
algorithms and is independent of $D$ for the echo-2 algorithm. Dynamics corresponds to pure squeezing, $N \approx 10^7$,
and time was chosen separately for each $D$ so that $F\approx0.99$.}
\end{figure}

Figure~\ref{fidelity_100D} displays fidelity in a $100$-dimensional system of
kicked rotators. It shows that echo-2 converges with several orders of
magnitude fewer trajectories than the echo-1, echo-1', and echo-N-1
algorithms. Figures~\ref{stat_err_PD} and~\ref{stat_err_PS} confirm that
$\sigma_{\text{echo-2}}$ is independent of $D$ while $\sigma_{\text{echo-1}}$,
$\sigma_{\text{echo-1'}}$, and $\sigma_{\text{echo-N-1}}$ grow exponentially
with $D$. The normalized echo-N-1 algorithm is the most efficient among the
methods with $M=1$.

To conclude, we have shown that not only QF, but also CF algorithms can be
unfeasible in complex systems due to the exponential scaling with
dimensionality. We have proposed an efficient CF algorithm for which this
exponential scaling disappears. In the special case of initial densities given
by characteristic functions all echo-$M$ and echo-N-$M$ algorithms (for $M>0$)
collapse into a single algorithm. In particular, the \textquotedblleft
natural\textquotedblright\ algorithm sampling from $\rho$ is equivalent to our
algorithm sampling from $\rho^{2}$. This may explain why high-dimensional
calculations were previously done only with characteristic functions. These
results should be also useful in applications computing more general overlaps of
phase space distributions. Finally, we have described a technique to analyze
efficiency of general trajectory-based algorithms. This can be useful in
developing approximate methods for QD of large systems. Our research was
supported by Swiss NSF with grants No. 200021\_124936 and NCCR MUST, and by EPFL.

\bibliographystyle{apsrev4-1}
%

\end{document}